\documentclass[preprint,review,12pt]{elsarticle}

\usepackage{amsmath}
\usepackage{soul}
\usepackage{amssymb}

\usepackage{xcolor}

\newcommand{\VST}{$V_\mathrm{ST}$}
\newcommand{\VSB}{$V_\mathrm{SB}$}
\newcommand{\VET}{$V_\mathrm{h}$}
\newcommand{\VES}{$V_\mathrm{ES}$}

\usepackage{changes}
\setaddedmarkup{\textcolor{blue}{#1}}
\setdeletedmarkup{\textcolor{red}{\sout{#1}}}

\journal{Sensors and Actuators: Physical}
\begin{document}

\begin{frontmatter}



\title{Effect of Overheat and Direct Flow Loading on the MEMS Bistable Flow  Sensor}


\author[inst1]{Ivan Litvinov}
\author[inst1]{Dan Refaeli}
\author[inst1]{Alex Liberzon}
\author[inst1]{and Slava Krylov}
\affiliation[inst1]{organization={School of Mechanical Engineering},
            addressline={Tel Aviv University}, 
            city={Tel Aviv Jaffo},
            postcode={6997801}, 
            country={Israel}}


\begin{abstract}



We present the findings from an experimental study of a MEMS flow sensor in which an initially curved, double-clamped bistable microbeam is the primary sensing element. Our research explores how the overheat ratio, direct flow loading, and turbulence-induced vibration affect the sequential snap-through (ST) buckling and snap-back (SB) release of an electrostatically actuated beam heated by an electric current. The sensor is fabricated from highly doped single-crystal silicon using a silicon-on-insulator (SOI) wafer. Positioned at the chip's edge, the microbeam is exposed to airflow, enabling concurrent dynamic response measurements with a laser Doppler vibrometer and a video camera.

Our research demonstrates that the overheat ratio can be significantly lower for this sensing principle than conventional thermal sensing elements, pointing to the potential for substantial energy savings. We also emphasize the significant impact of flow angles and vibrations on the critical ST and SB voltages, which are vital for the flow sensor's output. Additionally, we introduce the first direct experimental observation of the beam profile's time history during the snap-through /snap-back transition.

The potential impact of this research lies in developing more robust MEMS flow sensors with enhanced sensitivity and a better understanding of their response to environmental factors, which could have broader applications in fields such as aerospace, environmental monitoring, and industrial process control.

\end{abstract}

\begin{graphicalabstract}
\includegraphics[width=\textwidth]{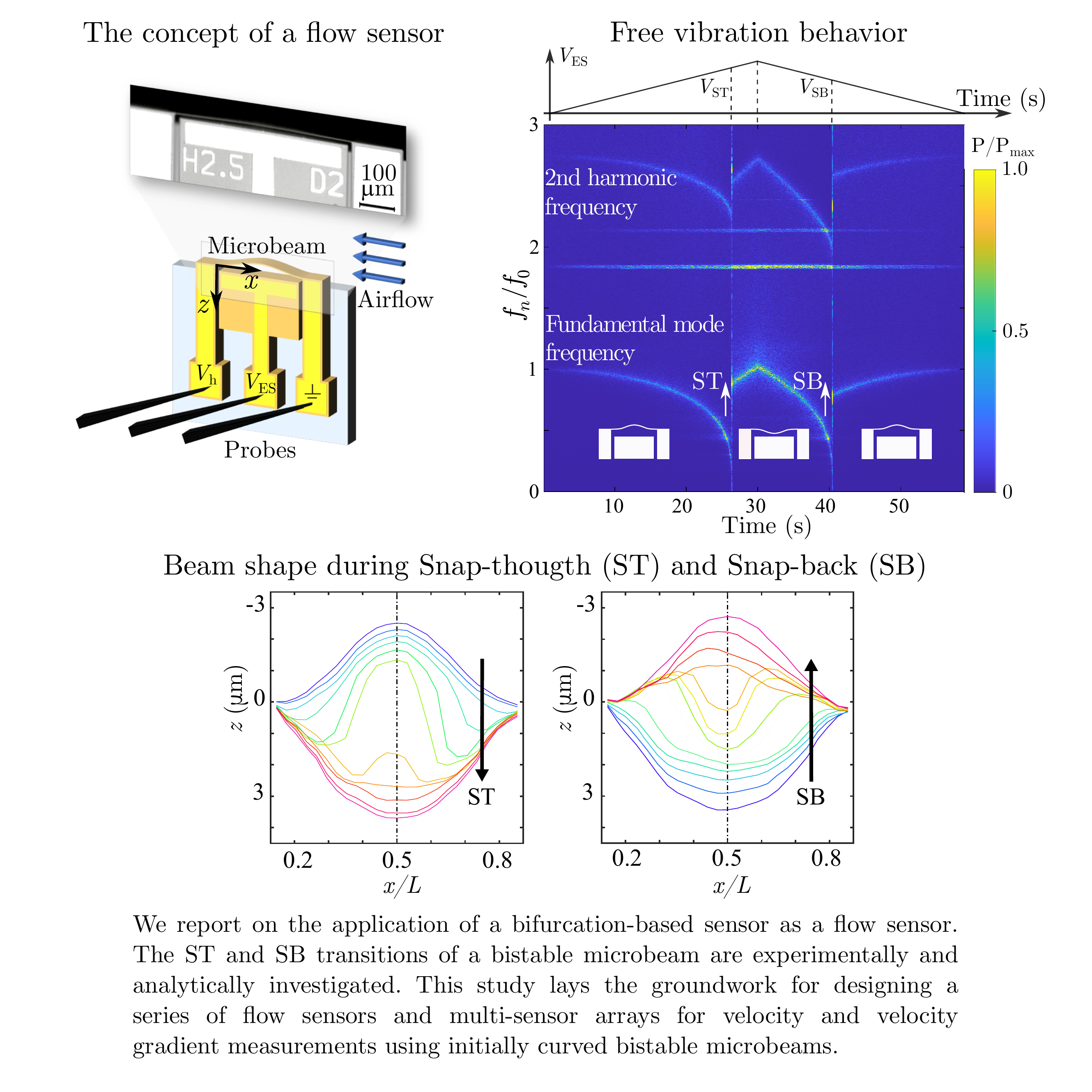}

\end{graphicalabstract}

\begin{highlights}
\item A MEMS bistable flow sensor based on the initially curved double-clamped microbeam located at the edge of the chip is presented
\item Influence of overheat ratio, direct flow loading, and turbulence-induced vibrations on the critical voltages are investigated
\item Direct beam profile measurement during the dynamic snap-through and snap-back transitions is presented.
\end{highlights}


\end{frontmatter}

\clearpage
\section{Introduction}
\label{sec:intro}
Microscale flow sensors play a key role in advancing scientific and technological endeavors. Among the various types of flow sensors, those based on microelectromechanical systems (MEMS) have become increasingly important for monitoring gas flow in industrial environments, sensing environmental flows, detecting marine conditions, and analyzing biomedical flows~\cite{ho1998micro}. 
These sensors have a wide range of applications in technical fields, such as ``smart skin'', the Internet of Things (IoT), and flight control (e.g.,~\cite{dong2017mems,shi2020progress}).

The utilization of MEMS-based flow sensors presents several advantages, including compactness, power efficiency, high sensitivity, and compatibility with mass production requirements. Consequently, a multitude of MEMS flow sensing methods have emerged over the past three decades~\cite{wang2009mems,kuo2012micromachined,ejeian2019design}. These sensors operate based on thermal or mechanical principles, and most previously developed flow sensors are based on the thermal principle~\cite{ejeian2019design}. Thermal flow sensors offer robust and straightforward manufacturing processes, linear sensor characteristics, and lack of moving parts, distinguishing them from mechanical flow sensing methods. However, the main drawbacks of the thermal sensors include high power consumption, calibration requirements, unwanted heat dissipation, and limitations in miniaturizing heat sources. The reduction in the size of the sensing element was the subject of extensive research endeavors ~\cite{le2021fabrication}. Recent advances in the design and fabrication of microelectronic devices have focused on reducing the power consumption of thermal flow sensors~\cite{xu2021cmos, yang2022foundry}. However, mechanical flow sensors continue to play a vital role in water flow, especially in laminar flow. For example, micro/nanocantilever-based sensors have found various applications in chemical and biological sensing~\cite{mathew2018review}. Mechanical flow sensors often rely on drag or lift force sensing~\cite{kalvesten1996integrated, svedin1998lift,svedin2003new}, frequency monitoring~\cite{bouwstra1990resonating}, or artificial hair sensor designs~\cite{dagamseh2011bio}. While mechanical flow sensors are sensitive enough to detect low flow rates, they face challenges in measuring higher flow rates. Although various architectures of MEMS flow sensors are being developed in both industry and academia, specific sensing capabilities are sought to meet the technical requirements of each application, depending on environmental and flow conditions.


Previous studies~\cite{krylov2018IEEE, kessler2018JMEMS} introduced an alternative flow sensing principle that combines thermal and mechanical responses in a bifurcation-type sensor. These studies showed that the electrostatically and electrothermally (Joule heated) actuated initially curved double-clamped silicon microbeam exhibits bistable behavior, in other words, remaining in one of the two equilibrium position states under the same load. 
As a result of this bistability, the curved beam shows high responsiveness to external stimuli near two equilibrium states 
and has excellent potential for sensor implementation. This property of bistable systems plays an instrumental role in many MEMS/NEMS-based devices~\cite{qiu2004curved, harne2013review}. When the mechanical structure is pushed to its stability limits, it becomes highly sensitive to even the slightest changes in heat transfer caused by flow. Reducing the overheating ratio (which is the ratio between the absolute temperature to which the microbeam is heated and the ambient temperature) can significantly decrease power consumption. Direct comparison of different sensors is challenging due to their completely different operating principles. However, note that the measured sensitivity of a bifurcation-type sensor based on the bistable curved silicon microbeam (1.9 V/m/s~\cite{krylov2018IEEE}) is comparable to or better than the state-of-the-art values reported for thermal flow sensors~\cite{Kuo2012}.

The feasibility of employing such devices for airflow sensing has been confirmed through theoretical analysis and experimental testing~\cite{krylov2018IEEE, kessler2018JMEMS,kessler2020flow,kessler2021sampling}. However, enhancing our understanding of the device's response to various influencing factors is imperative to effectively utilize the laboratory prototype in airflow measurements within practical scenarios. These factors include overheat ratios, the sensitivity of the microbeam to direct flow loading, and noise and vibrations that may arise during turbulent flow measurements, such as those over aircraft wings or wind turbine blades.

This study aims to investigate how the sensor interacts with these factors. An especially important result is the direct visualization of the beam's shape during transitions between two equilibrium states. We added a multi-point Doppler laser vibrometer to the diagnostic system introduced in \cite{kessler2020flow} to achieve this. The outcome of this extension has revealed a previously theoretically predicted phenomenon – asymmetric snap-through  visualization of an electrostatically actuated beam, as described in \cite{medina2012symmetry}. Furthermore, this finding sheds light on the limitations of the mathematical model presented in the paper, highlighting the need for further refinements.


\section{Materials and methods}\label{sec:methods}

\subsection{Sensor and chip design}
\begin{figure}[!ht]
\centering
\includegraphics[width=\textwidth]{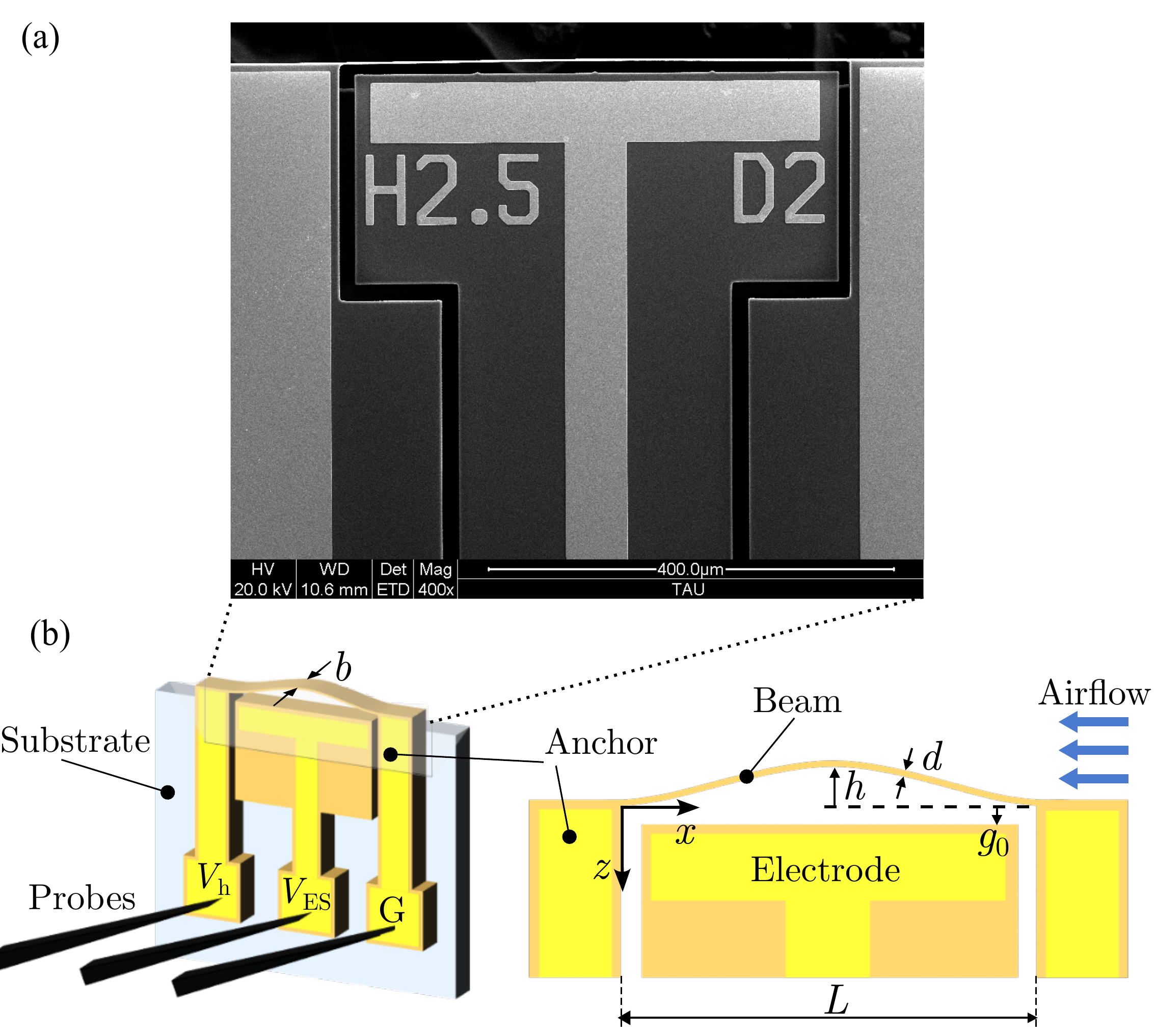}
\caption{(a) A scanning electron microscope micrograph of the device, and (b) Schematic view of the beam, its main geometric parameters, and the airflow direction (not to scale). The double-clamped curved beam is shown in the initial upward curved configuration. The values \VET~and \VES~are the heating and electrostatic voltages, respectively (G is a ground).}  
\label{fig:Schematics} 
\end{figure}

The bistable flow sensor based on an initially curved microbeam was introduced in ~\cite{kessler2020flow,kessler2021sampling}. We briefly review the device design and its operational principle for clarity and completeness. We use an initially curved bell-shaped electrostatically actuated double-clamped beam, as shown in Figure~\ref{fig:Schematics}(a), as the sensing element. The beam is fabricated from the device layer of a silicon-on-insulator (SOI) wafer using the SOIMUMPs\textsuperscript{TM} process~\cite{soimumps}. The beam, anchors, and electrode are all highly doped single crystal silicon with a resistivity of 0.02 $\Omega \cdot$cm. The beam is designed to perform an in-plane motion parallel to the substrate and to deflect in the $z$ direction. The nominal dimensions of the beam ``as designed'' dimensions are given in Table~\ref{tab:dimensions}, and their definitions are shown in Figure~\ref{fig:Schematics}(b). 

\begin{table}[!ht]
\small
\centering\caption{The nominal, as designed, dimensions of the beam used in the experiments.}
\label{tab:dimensions}      
\begin{tabular}{lc}
\hline
Parameter & Size ($\mu$m)    \\ \hline

Length, $L$  & $500$  \\
Width, $b$ & $25$  \\ 
Thickness, $d$ & $2$  \\ 
Initial elevation, $h$  & $2.5$  \\
Beam-electrode distance, $g_0$  & $10$ \\ 
\hline
\end{tabular}
\end{table}

The operational principle of the device relies on coupled thermal and mechanical sensing. It begins with an initially curved bistable microbeam, subjected to controlled Joule heating, using a weak electric current regulated by a constant heating voltage, denoted as \VET. The voltage difference \VET~is applied across the two beam anchors, with one of them serving as the ground (G). Simultaneously, airflow removes heat from the microbeam, as illustrated in Figure~\ref{fig:Schematics}(b). 

The electrostatic time-periodic force acting on the beam is controlled by the actuating voltage \VES, applied to an electrode, as marked in Figure~\ref{fig:Schematics}. Increasing \VES~beyond a critical snap-through (ST) voltage, denoted as \VST, triggers a dynamic shift from the microbeam's first to the second stable equilibrium position. On the contrary, reducing the voltage below the lower critical value \VSB, causes the beam to snap back (SB). The interaction between Joule heating and airflow cooling influences the beam's curvature and the critical voltage values. Measurements of \VST~and \VSB enable an estimate of air velocity. The enhanced performance is due to the high sensitivity of the bistable beam near the ST and SB points~\cite{krylov2018IEEE, kessler2018JMEMS}. Additionally, the device in its present design possesses a unique capability to measure the flow velocity at two closely spaced points corresponding to two stable positions of the microbeam, thus allowing for the direct measurement of the flow velocity gradients, as discussed in~\cite{kessler2020flow}.


\subsection{Setup}\label{sec:setup}

\begin{figure}[!ht]
\centering
\includegraphics[width=\textwidth]{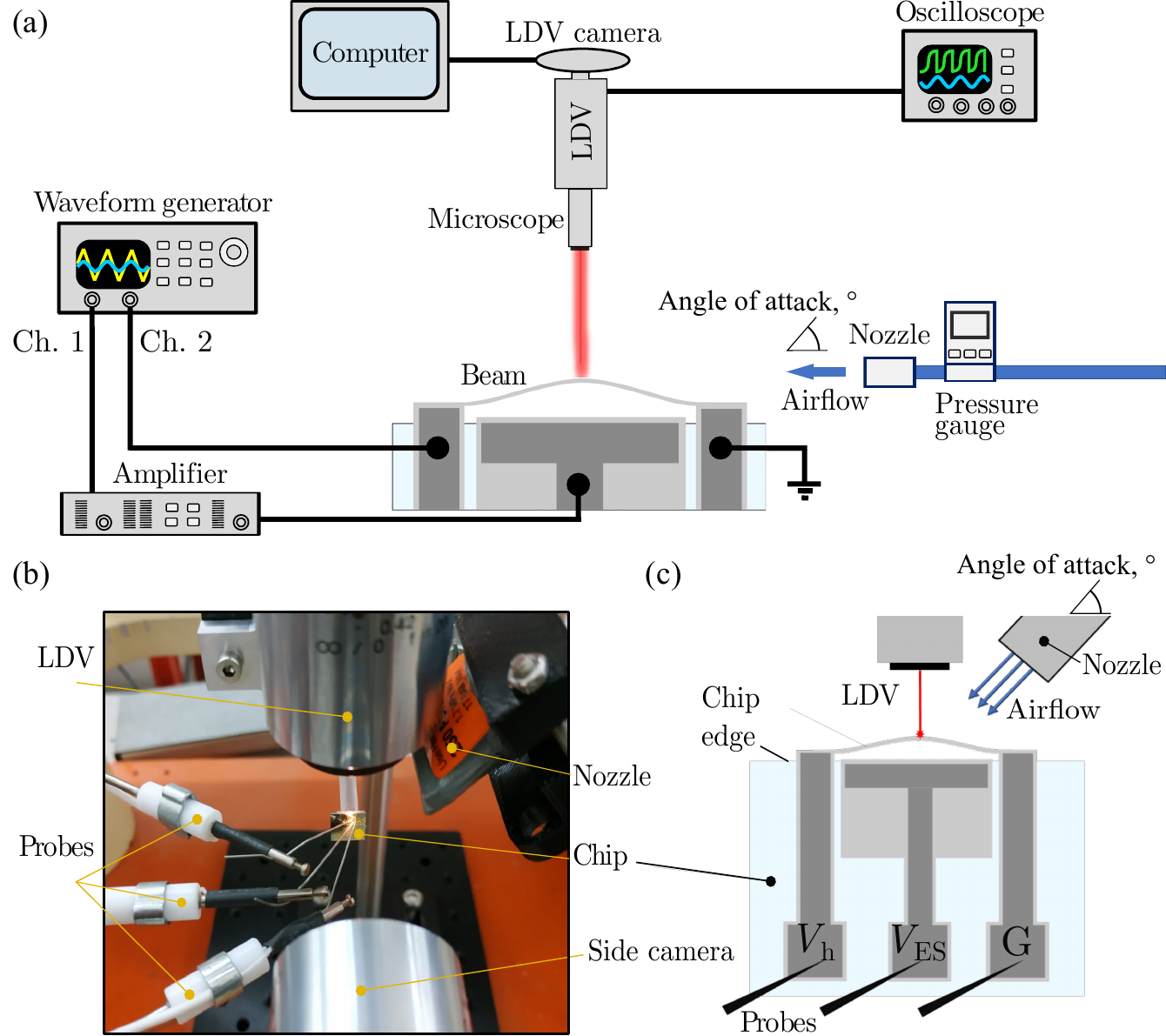}
\caption{(a) Overview of the experimental setup. (b) Optical image of the experimental setup and (c) schematics of the chip from the perspective of the side camera, with the laser from the LDV, and the airflow at an angle of attack ($^{\circ}$) relative to the beam.
\label{fig:setup} }
\end{figure}

The experimental setup is illustrated in Figure~\ref{fig:setup}(a). During the experiments, an AFG3022C waveform generator from Tektronix is employed. It provides a voltage through channel 1, subsequently amplified through a signal amplifier (Trek PZD350A). This amplified voltage is then used to apply \VES~ to the electrode. Furthermore, to generate the electric current required for Joule heating within the beam, a constant heating voltage \VET~is applied across the beam's anchors, with this voltage being supplied by the same waveform generator through channel 2.

The real-time displacements of the beam's center are recorded using a single-point laser Doppler vibrometer (LDV, Polytec, OFV534). The LDV focuses its laser beam on the side face of the beam, which has a width of $b=25\;\mu$m. The values of \VST~and \VSB~are extracted from the time series recorded with an oscilloscope (DSOX2004A, Keysight) synchronously with the LDV signal.

To regulate the flow rate, we employ a pressure gauge (P-30PSIG-D/5P, Alicat) to monitor the pressure in the air supply system. A rectangular nozzle (1"~High Power, EXAIR) generates a wide, uniform jet airflow toward the microbeam. The air supply is calibrated using a commercial anemometer (CTV 110, Kimo). Since the spatial resolution of the anemometer cannot ensure an accurate measurement of the uniform velocity distribution within the jet core, especially when the sensor is placed at a distance from it, we have experimentally estimated the uncertainty of the sensor position with respect to the flow nozzle and consequently the uncertainty of the velocity near the sensor. We estimate the maximum uncertainty of the velocity measurement near the sensor to be $\pm$2.8 m/s mainly to demonstrate the feasibility and working principle of the velocity sensor.
Our experiments cover a range of air velocities ranging from 0 to 15 m/s.  Additionally, it is worth noting that the airflow can be applied at various angles of attack, shown in Figure~\ref{fig:setup}.


The beam is strategically placed at the edge of the chip, as depicted in Figure~\ref{fig:setup}(b,c). This positioning offers the advantage of simultaneously capturing real-time beam velocity from a top view through LDV, while also enabling direct imaging of ST and SB events using a custom-built optical arrangement. The optical setup comprises a video camera (UI-6250SE-M-GL, IDS) and a horizontally mounted tube microscope (Navitar). In most of our experiments, the side camera's use is primarily limited to the initial setup preparations, as the bulk of quantitative measurements are conducted through the LDV.

\subsection{Measurement methodology}
\label{sec:exp_methods}

\begin{figure*}[!ht]
\centering
\includegraphics[width=1\textwidth]{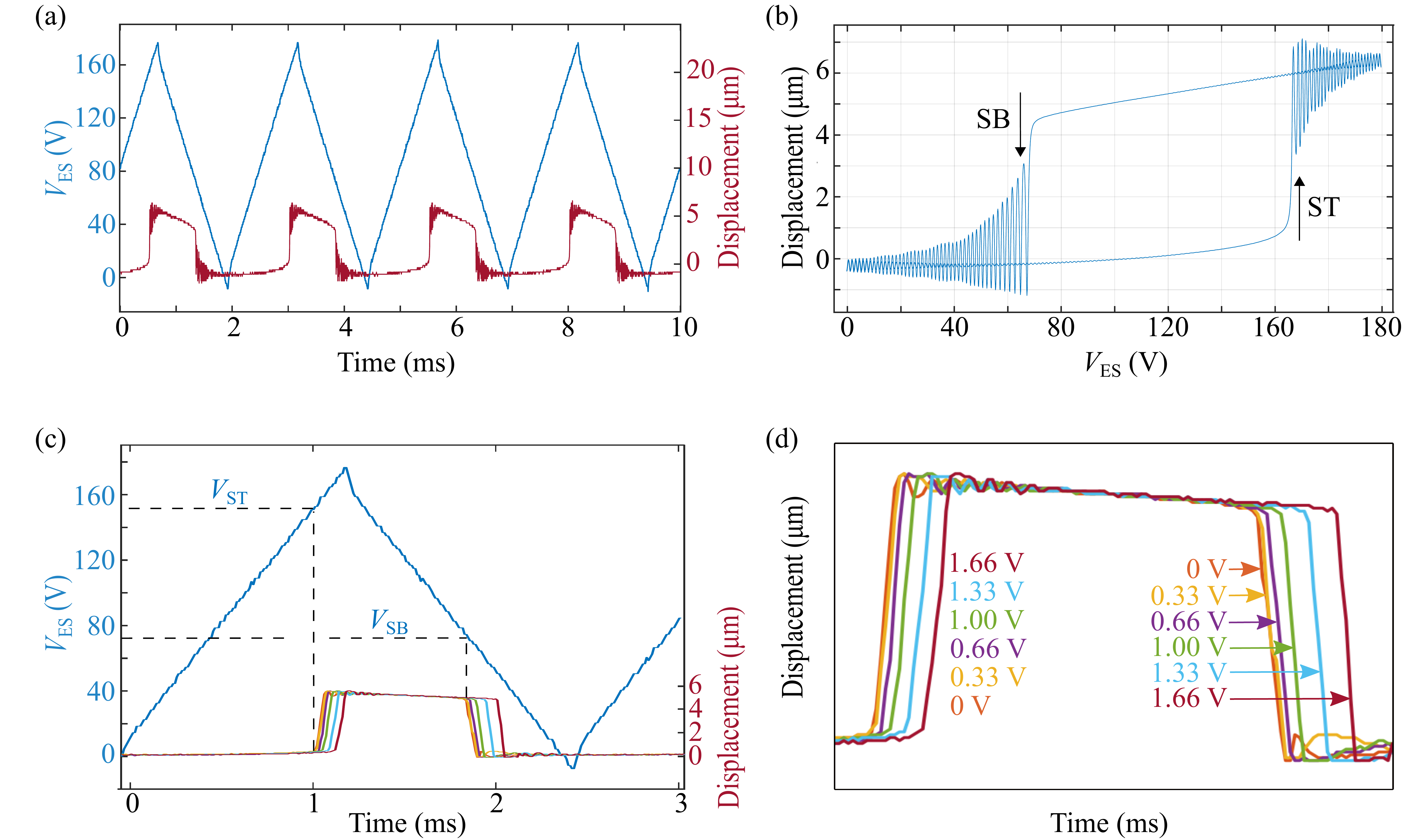}
\caption{
(a) Cyclic time history plot of the beam displacement (red) versus the voltage \VES~(blue) between the beam and the electrode (\VET=0 V).
(b) Displacement as a function of the actuating voltage on the beam.
(c) Electrostatic voltage (blue) and beam displacement (the other colors) as a function of time. In contrast to (a) and (b), a track filter is activated here to filter out fluctuations. 
(d) A close-up of the beam displacement from six different heating levels \VET=0, 0.33, 0.66, 1, 1.33, and 1.66 V.
\label{fig:Timehistory}  }
\end{figure*}

Figure~\ref{fig:Timehistory}(a) illustrates the time history of a cyclic voltage \VES~applied to the electrode (labeled in blue, left vertical axis) and the corresponding $z$ value of the beam midpoint ($x=L/2$), which is obtained using LDV (labeled in red, right vertical axis). The electrostatic voltage signal \VES~is a linear ramp with a constant rate of about 170~V per period. 
The analog output from the LDV controller and the input voltage signal are looped back into a digital storage oscilloscope, allowing for the synchronized recording of their time histories. The determination of \VST~and \VSB~values occurs during the post-processing phase. This is achieved by identifying the displacement threshold transition events within the recorded data, as illustrated in Figure~\ref{fig:Timehistory}(b). These threshold events correspond to a 5 $\mu$m jump in displacement, serving as triggers for ST and SB events.

The actuation process is repeated for four cycles, operating the beam at 400 Hz. Averaging is used to mitigate the variability of the data, facilitating the calculation of the mean value and its associated standard error. It is important to note that the mechanical response can be considered quasi-static due to the significantly lower frequency of the chosen electrical signal, which is on the order of $\mathcal{O}(10^5)$~Hz compared to the fundamental frequency of the beam.




The time history of the beam center displacement is presented in Figure~\ref{fig:Timehistory}(c) for different \VET~voltage levels. Crossing of dashed lines with the electrostatic voltage linear ramp signal helps to indicate the values of \VST~ and \VSB~, at the moments of transitions, for the case of the heating voltage \VET~equal to 0 V. A closer examination of the zoom in view in Figure~\ref{fig:Timehistory}(d) reveals a noticeable shift in the time when the threshold is crossed for increasing heating voltage \VET~in the range 0$\sim$1.66~V. This shift results in distinctly different critical electrostatic voltages \VST~and \VSB. We will address this shift quantitatively in the following, but we first address the bias in these measurements due to the LDV heating of the beam.

\subsection{LDV-Induced Beam Heating Bias}


The specific setup explained above enables us to obtain unique results, including directly measuring the beam response time history during the transition using LDV. However, the LDV system has a drawback known as the ``loading effect,'' which inadvertently heats the beam during measurements, as discussed in \cite{kassie2020effect}. We employ digital camera imaging of the beam to assess the bias on the measured \VST~ and \VSB~ values and compare the results obtained with and without LDV. It is important to note that the image processing method for detecting ST and SB events is less precise than the LDV approach. Therefore, we extended the period to 60 seconds in these experiments to achieve more accurate results. This extension was necessary because of the lower temporal resolution of the video camera, which demanded a longer measurement duration for precise observations. Consequently, the bias due to LDV heating could be more prominent than in our experiments' case of high-frequency measurements due to a significantly longer measurement period.

%



\begin{figure}[h!]
\centering
\includegraphics[width=0.9\textwidth]{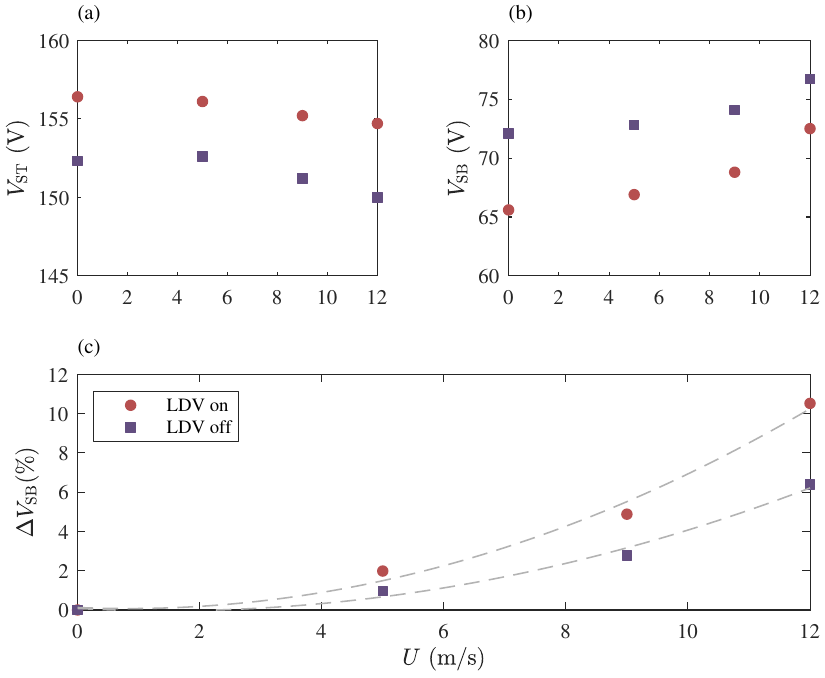}
\caption{(a) Critical snap-through voltage (\VST) and (b) critical snap-back voltage (\VSB) as a function of the flow velocity $U$. Circle markers indicate measurements with LDV, while square markers indicate measurements without LDV. (c) Sensitivity of the snap-back voltage \VSB, measured as a relative voltage change at increasing flow velocity, normalized by the reference value at $U=0$ m/s. The dashed lines represent a quadratic polynomial fit. 
}
\label{fig:LDV0}
\end{figure}

The figures in Figure~\ref{fig:LDV0}(a, b) show the results with ``LDV on'' and ``LDV off,'' represented by circles and squares.

The LDV system in the experiment causes an increase in \VST~and \VSB of approximately 4 V and 6 V, respectively. To further quantify the influence of LDV heating, we examine the relative change in SB voltage, denoted $\Delta V_{\mathrm{SB}}=100\times(V_{\mathrm{SB}}-V_{\mathrm{SB}}|{U=0})/V_{\mathrm{SB}}|{U=0}$ in Figure~\ref{fig:LDV0}(c). Notably, $\Delta V_{\mathrm{SB}}$ increases from 6\% to 10\% at an airflow velocity of 12 m/s when the LDV is active. We will discuss this bias further in Section~\ref{sec:heat}, where we discuss the behavior of the flow velocity sensor due to Joule heating.

\subsection{Analytical model}
\label{model}

\subsubsection{Beam equilibrium}

To understand the sensor behavior and interpret our experimental findings, we explored a reduced order model of the double-clamped heated microbeam with initial curvature, as shown in~Figure~\ref{fig:Schematics}. Although previous models in~\cite{medina2014experimental, krylov2018IEEE} addressed specific aspects of sensor phenomena, such as aerodynamic loading, overheat ratio, air cooling, or dynamic effects, each model focused on different aspects. In this work, we present a comprehensive model that describes the combined influence of all these effects and their interactions. 

The motion of the beam subjected to an axial force $N$ (positive in compression) and a distributed actuating force $f_a$ can be described by the following equation Eq.~\eqref{eq:BeamGeneral}. 

\begin{equation}
\begin{split}
\rho A \ddot{z}\!+\!c\dot{z} + EI_{yy}\left(z^{''''}-z_0^{''''}\right) + \left[N - \frac{EA}{2L}\int_0^L \left(z'^2 - z_0'^2\right)dx\right]z'' = f_a(z), \\
 z(t,0)=z(t,L)=0, \; z(0,x)=z_0(x)
\label{eq:BeamGeneral}
\end{split}
\end{equation}

Here, $z(x,t)$ and $z_0(x)$ are the deformed and the initial (as-designed) elevations of the beam above its anchored ends, respectively (see Figure~\ref{fig:Schematics}(b)); $EI_{yy}$ and $EA$ are the bending and axial stiffness of the beam, where $I_{yy}={bd^3}/{12}$ is the second moment of area of the cross-section, $A=bd$ is the cross-sectional area of the beam, $c$ is the viscous damping coefficient, $E$ is Young's modulus, $\rho$ is a mass density and $\dot{(\;)}$, $(\;)'$ denote the time and the axial coordinate $x$ derivatives, respectively.

The axial force: 
\begin{equation}
 N=\sigma_\text{res}A+\frac{\alpha EA}{L}\int_0^{ {L}}\theta d {x}=\sigma_\text{res}A+\alpha EA\overline{\theta}
 \label{eq:axial_force}
 \end{equation}
%
\noindent is generated by both the beam heating of the beam \cite{gerson2010electrothermal} and by a residual stress $\sigma_{res}$, estimated to be a few MPa \cite{krylov2018IEEE,kessler2018JMEMS}.  In Eq.~(\ref{eq:axial_force}), $\alpha$ is the thermal expansion coefficient of Si, and $\overline{\theta}$ is the length-averaged difference between the beam temperature $T( {x})$ and the ambient temperature $T_\infty$. In our case, we use Joule heating by applying a voltage \VET~between the ends of the beam to induce axial loading.

The beam is actuated by an electrostatic force, which is parameterized by voltage \VES, and can be described by the following equation
%
\begin{equation}
 f_a(z)=\frac{\varepsilon_0 b V_{\text{ES}}^2}{2(g_0-z)^2}\left(1+0.65\,\frac{\langle g_0-z \rangle}{b}\right),
 \label{eq:ES}
 \end{equation}
\noindent where $\varepsilon_0$ is the vacuum permittivity and $\langle g_0-z 
 \rangle = \int_{0}^{L} (g_0-z) \,dx/L$ is the average distance between the beam and the electrode. Here, we use a first-order fringing field correction \cite{osterberg1997m} because the term $\langle g_0-z_0 \rangle /b \approx 0.4$ is not negligibly small.

We construct the single-degree-of-freedom reduced-order model associated with Eq.~\eqref{eq:BeamGeneral} using the Galerkin decomposition based on the following approximations of $z(x,t)$ and $z_0(x)$:
\begin{equation}
 z(x,t)= q(t) \phi(x),\quad z_0(x)=- {h}\phi( {x}).\; 
\label{eq:base_function}
\end{equation}
\noindent where \(\phi( {x})=1/2[1-\cos(2\pi x/L)]\) is the base function, \(q\) and \(h\) are the deformed and the initial midpoint elevations of the beam, respectively. Note that $h$ is positive, and the minus sign is due to the direction of the positive $z$-axis defined in Figure~\ref{fig:Schematics}(b). 
%
Substituting the representation in Eq.~\eqref{eq:base_function} into Eq.~\eqref{eq:BeamGeneral}, multiplying the resulting expression by $\phi(x)$, integrating by parts, and using the orthogonality of base functions in Galerkin's formalism, we obtain the following equation expressed with non-dimensional variables. 
%
\begin{eqnarray}
\ddot{w}_m+\frac{\dot{w}_m}{Q} + w_{m} \left( 1-\!{n}+\frac{\eta^2 }{8}\right)-
\frac{3w_m^2\eta }{16}   + \frac{w_m^3}{16}=\!
-{n\eta}+\frac{\beta}{\sqrt{(\gamma+\eta-w_m)^3}}.\;\;
\label{EQ:beamRO}
\end{eqnarray}
%
%
%
%
Non-dimensional quantities used in the reduced order model are summarized in Table~\ref{tab:nondim}.
%
\begin{table}[!ht]
\small
    \centering\caption{Quantities used in the development of the reduced-order model. \label{tab:nondim}}
    \begin{tabular}{ll}
    \hline
       $r=\sqrt{I_{yy}/A}$  & Gyration radius \\
       $\gamma = g_0/r$ & Normalized beam-electrode distance\\
       $\eta = h/r$ & Initial midpoint elevation\\
       $w_m = \eta+{q}/r$  & Midpoint deflection \\
       $N_e = 4\pi^2 EI_{yy}/L^2$  & Euler's buckling force \\
       $n=\frac{N}{N_e} = (\sigma_\text{res}A+\alpha EA\overline{\theta})/N_e$   & Normalized axial force\\
        $\beta=\frac{\varepsilon_0bL^4V_{\text{ES}}^2}{8\pi^4r^3\sqrt{\gamma}EI_{yy}}(1+0.65\,\frac{\langle g_0-z \rangle}{b})$ & Voltage parameter\\
        $Q=4\pi^2/(c {L}^2)\sqrt{EI_{yy}\rho A/3}$ & Quality factor\\
        \hline
        
    \end{tabular}
\end{table}
For the case of quasi-static actuation, we can omit the terms containing time derivatives, then Eq.~(\ref{EQ:beamRO}) could be solved for $w_m(V_{\mathrm{ES}})$.

%
%


Equation~\eqref{EQ:beamRO} highlights the flow sensor operational scenario: flow reduces the average temperature difference between the beam and the air, $\overline{\theta}$, affecting the axial force ($n$ in Table~\ref{tab:nondim}) and altering the critical ST and SB voltages expressed through $\beta$. To predict the \VST~and \VSB~values and the flow, it is necessary to quantify the convective heat transfer and relate $\overline{\theta}$ to flow velocity over the beam.

For simplicity of analytical prediction, we limit our analysis to steady-state heat transfer and neglect radiation. In this simplified case, the heat transfer equation in the beam is governed by the equation~\eqref{eq:heat_dim}:
\begin{equation}
-(\kappa  {\theta}')'={J}^{2}\rho_{{e}} -\frac { P\tilde{h}}{A}{\theta}  \;,\; \theta(0) = \theta(L) = 0,
\label{eq:heat_dim}
\end{equation}
\noindent where $\kappa$ is the thermal conductivity, $\tilde{h}$ is the average convective heat transfer coefficient, and $P=b+2d$ is the perimeter of the beam's cross-section. $J=I/A=V_{\mathrm{h}}/(\rho _{e}L)$ and $\rho _{e}$ represent the electric current density and the resistivity of the beam's material, respectively. Assuming that $\kappa$ and $\rho_e$ are temperature-independent around the operating point, we can find the averaged solution of Eq.~(\ref{eq:heat_dim}) as follows,
\begin{equation}
\overline{\theta}=\frac{V_{\text{h}}^2}{4\kappa \rho_e m^2}\left(\! 1-\!\frac{\tanh(m)}{m}\right), \;\; m=\frac{
{L}}{2}\sqrt{\frac{P\tilde{h}}{\kappa A}}.
\label{eq:thermal_solution}
\end{equation}
%


In Eq.~\eqref{EQ:beamRO}, $\overline{\theta}$ is given by Eq.~(\ref{eq:thermal_solution}), and $\tilde{h}$ can be estimated using the convective heat transfer theory. Due to the small size of the beam, the relevant dimensionless parameter $Re = U L/\nu$ is small, and we can use the laminar boundary layer analysis. $\tilde{h}$ can be determined using the non-dimensional Nusselt number, $Nu$, as $\tilde{h} = Nu \, \kappa_\text{air}/L$. In our analytical model, we assume that heat exchange is characterized by a laminar boundary layer on a flat plate (microbeam) and utilize the laminar convection coefficient $Nu = 0.664 \, Pr^{1/3} \,Re^{1/2}$. Here, the nominal values for airflow can be used: Prandtl number, $Pr = 0.7$, the thermal conductivity of air ${\kappa_\text{air}=0.0257}$~W/(mK), and the kinematic viscosity $\nu= 1.5 \times 10^{-5}$ m$^2$/s ~\cite{krylov2018IEEE}. The material properties of single crystal silicon and the rest of the constants required for the application of the model are given in Table~\ref{tab:properties}.

\begin{table}[!ht]
\small
\centering\caption{Properties of single crystal silicon (correspond to $T=300$K).
\label{table1}}
\begin{tabular}{lc}
\hline
Parameter & Value \\ \hline
Young's modulus, $E$ & 169 GPa \\
Thermal expansion, $\alpha$ & $2.6 \times 10^{-6}\; 1/^{\circ}$C \\ 
Thermal conductivity, $\kappa$ & $150$ W/(mK) \\ 
Vacuum permittivity, $\varepsilon_0$ & $8.85 \times 10^{-12}$ F/m \\ 
Resistivity, $\rho_{e}$ &  $0.02\;\Omega$ cm \\
Density, $\rho$ &  $2.33$ g/cm$^3$ \\ 
Residual stress, $\sigma_{res}$ & -$2.7$  MPa\\
\hline
\end{tabular}%
\label{tab:properties}
\end{table}

In this study, we revisit the conclusions of the works~\cite{kessler2018JMEMS, krylov2018IEEE,kessler2020flow,kessler2021sampling}  and perform detailed experimental investigations of the influence of several additional, not addressed in the previous works, effects on the sensor. 
%


\subsubsection{Frequency response}

Besides the quasi-static analysis, we also investigate the beam's frequency response, which involves exploring the effects of beam oscillations after ST and SB transitions. This encompasses studying the fundamental frequency of the free vibrations, $f_\text{res}$, corresponding to the linearized version of Eq.~\eqref{EQ:beamRO}:

%
\begin{equation}
f_\text{res}=\frac{\omega_0}{2\pi}\sqrt{1-n +\frac{\left( 3w_m^2 + 2\eta^2 - 6 \eta w_m  \right)}{16}  -\frac{3\beta}{2\sqrt{(\gamma+\eta-w_m)^5}}},
\label{eq:NF}
\end{equation}
\noindent where $\omega_0=4\pi^2\sqrt{EI_{yy}/(3\rho A L^4)}$ is the fundamental frequency approximation of the initially straight unforced beam and $w_m(\beta,\overline{\theta})$ is the solution of the equilibrium equation Eq.~\eqref{EQ:beamRO}. More information on this derivation can be found in~\cite{krylov2010dynamic,krakover2016displacement,kessler2021sampling}.

\section{Results and discussion}

In this section, we outline the results of measurements revealing various factors on sensor behavior during operation, such as the overheat ratio, flow loading, noise, and vibrations. Finally, for the first time, we demonstrate the real-time beam shape during ST-SB transitions.

\subsection{Joule heating effect}
\label{sec:heat}

\begin{figure}[!ht]
\centering
\includegraphics[width=\textwidth]{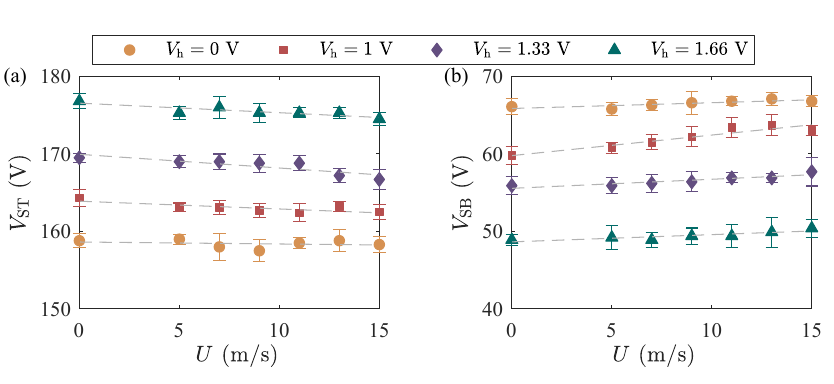}
\caption{(a) Critical snap-through voltage (\VST) and (b) critical snap-back voltage (\VSB) as a function of the flow velocity $U$ at different \VET~levels. Dashed lines emphasize linear trend lines. 
\label{fig:Joule's_heating}
}
\end{figure}

We first present the quantitative results of various runs conducted at different Joule heating voltages (\VET=0, 1, 1.33, and 1.66 V). In Figure~\ref{fig:Joule's_heating}(a, b), we illustrate the variations in \VST~and \VSB~in response to Joule heating at different \VET~values across a flow velocity range of 0 to approximately 15 m/s. Our results demonstrate that Joule heating causes a consistent offset in all curves, and we achieved high sensitivity across all voltage levels, highlighting the sensor's performance robustness. Note that for all these measurements,  we used LDV with the inherent bias in  \VST~and \VSB~due to laser heating of the beam. 

To better understand the flow sensor's overheat ratio, we studied the beam's response to various electrostatic voltages that affect Joule heating. (\VES). We employ an analytical model, detailed in Sect.~\ref{model}, to systematically compare the model's predictions and the experimental results. This in-depth analysis allows us to assess how varying electrostatic voltages influence the sensor's behavior, shedding light on its performance in different operational scenarios.

\begin{figure}[!ht]
\centering
\includegraphics[width=\textwidth]{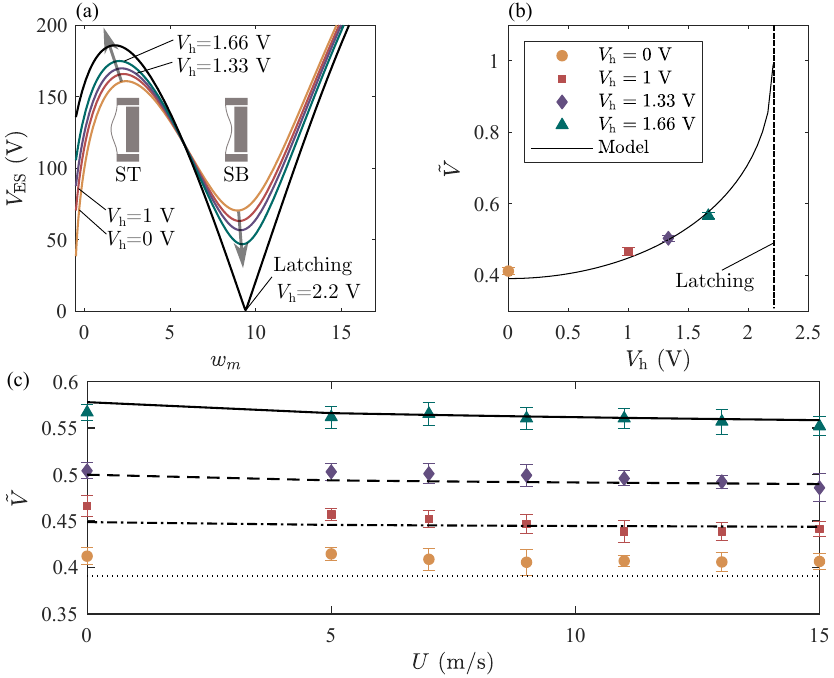}
\caption{(a) Modeled beam response to the electrostatic voltage \VES~at various heating voltages \VET, with a few curves shown for clarity. Insets show the beam's convex (before ST) and concave (after ST) configurations. 
(b) Comparison between experimental (symbols) and modeled (solid curve) voltage $\tilde{V}$ as a function of voltage \VET. Error bars represent a standard error of the mean.
(c) Critical voltage indicator ($\tilde{V}$) as a function of flow velocity $U$ at different \VET~levels. The model results are represented by lines (dotted, dash-dot, dashed, and solid for \VET=0, 1, 1.33, and 1.66 V, respectively). The error bars represent the standard error. }
\label{fig:model_response}
\end{figure}

Figure~\ref{fig:model_response}(a) shows the analytical predictions of changes in the electrostatic voltage \VES~as a function of the normalized midpoint displacement $w_m$~of the beam.
The other limit point (local minimum, SB) on the curve corresponds to the SB voltage $V_{\text{SB}}$~at which the beam returns to its original state. It is important to note that bistability can manifest even without additional beam heating ($V_{\text{h}}=0$), attributed to the beam's initial curvature.

As \VET~increases, $V_{\text{ST}}$ increases, while the 
$V_{\text{SB}}$ decreases. Once a certain level of heating is reached, $V_{\text{SB}}$ reaches zero. The meaning is that the beam can remain in the switched configuration at zero actuating voltage. At even higher  \VET~, a force in the opposite (negative $z$) direction should be applied to switch the beam back. This behavior is
known as the latching effect \cite{simitses2012dynamic}. 



To quantify the relationship between critical voltages $V_{\text{ST}}$ and $V_{\text{SB}}$, to both flow velocity and Joule heating, we introduce a universal critical voltage indicator denoted as $\tilde{V}$, calculated as ${\tilde{V}=(V_{\text{ST}}-V_{\text{SB}})/(V_{\text{ST}}+V_{\text{SB}})}$. This introduction of a universal indicator serves two key purposes. First, it consolidates the information into a single value, streamlining the analysis rather than dealing with separate values for ST and SB voltages. Secondly, it attempts to disentangle the influence of the overheating ratio effect from the impact of varying ambient temperatures, a factor that can fluctuate under different experimental conditions. The point is that the sum of the ST and SB voltages, at least within the single-degree-of-freedom model framework, is much less influenced by varying ambient temperatures. Specifically, following the single-degree-of-freedom model the sum $V_{\mathrm{ST}}+V_{\mathrm{SB}}=\mathrm{const}$, and is independent of the temperature (in the case of an actuation by a "mechanical", deflection-independent force). Therefore, the indicator is considered a convenient quantity for comparative studies.

Figure~\ref{fig:model_response}(b) compares the analytical model with the experimental results, illustrating the changes in $\tilde{V}$ as a function of \VET. The model predictions are consistent with the experimental observations. Furthermore, it predicts the value of the heating voltage value ($V_{\text{h}} = 2.2$~V) corresponding to latching, which was not reached in the experiment. It is seen that if the latching effect is achieved, the parameter $\tilde{V}$ equals one.

The Joule heating effect in terms of $\tilde{V}$ as a function of the flow velocity $U$ is shown in Figure~\ref{fig:model_response}(c). The figure also shows, represented by the different types of lines, the theoretical results obtained using Eq.~\eqref{EQ:beamRO}. 
There is no perfect agreement between the model and the experiment, especially for the low heating case (\VET=0 and 1 V). Note that the reduced-order model could not take into account the effect of velocity when the Joule heating is ``off'' (\VET=0 V), but there may be a small temperature difference between the airflow and the microbeam. Nevertheless, the reduced-order model accurately captures the trends in the critical voltage change, indicating that the primary factor in velocity measurement is the change in average beam temperature, as suggested by the model.  The graphs also show that all curves have nearly the same slope angle of 0.015 V/(m/s). 
This suggests that the beam overheat ratio is not a critical parameter for the sensor, which raises the possibility of significantly reducing energy consumption at the required sensitivity level.




\subsection{Flow angle sensitivity}

\begin{figure}[!ht]
\centering
\includegraphics[width=.9\textwidth]{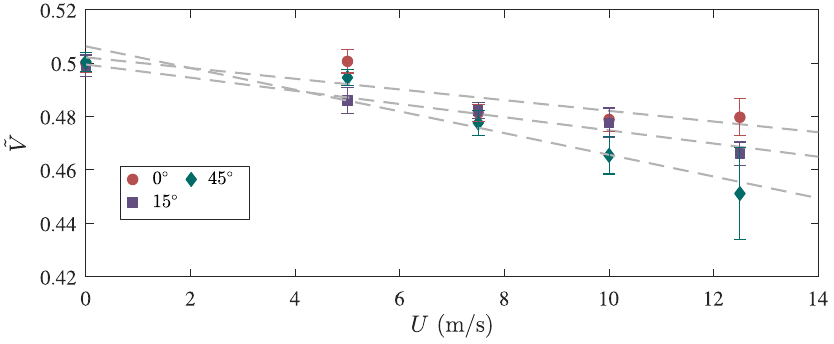}
\caption{Influence of flow direction by changing the angle of attack (\VET=1.33 V).} Dashed lines represent visual trends. Error bars represent a standard error of the mean.
\label{fig:ang_Joul} 
\end{figure}

This section delves into the influence of flow angle on the sensor's response. The sensor's angle sensitivity is evaluated in the plane parallel to the sensor, as depicted in Figure~\ref{fig:setup}(b, c), with the sensor's long side aligned with the flow direction. The results of this impact, illustrated in Figure~\ref{fig:ang_Joul}, reveal that the $\tilde{V}$ values remain consistent at low velocities. However, as flow velocities increase, deviations appear for different flow directions defined by the angles of attack~$0^{\circ},\ 15^{\circ}$, and $45^{\circ}$.

The effect of flow direction on the $\tilde{V}$ value is noticeable. Our sensor behaves similarly to miniature hot-wire sensors for measuring turbulent flows, as shown in reference~\cite{borisenkov2015multiarray}. Due to its more straightforward manufacturing process and reliable angular response, the bistable microbeam has potential in future multi-array flow sensors capable of measuring all three flow velocity components.

\subsection{Frequency response and sensitivity to vibrations}

In Figure~\ref{fig:LDV0}, it is evident that the sensors respond to flow velocity even when no LDV or Joule heating is involved. This suggests an additional mechanism associated with airflow in our experiment. Since the flow runs parallel to the flow sensor, this effect is unlikely to be attributed to mean flow drag or lift forces, as these forces are negligible in this particular configuration.

We considered that the device might be sensitive to turbulent velocity fluctuations or chip vibrations. To investigate this, we initially determined the natural frequency of the vibration of the beam around various equilibrium configurations using LDV, following a similar approach to our previous study~\cite{kessler2021sampling}. 


To achieve this, we introduced time-harmonic (AC) components to the steady (DC) electrostatic actuating voltage, resulting in the equation $V_{\text{ES}}(t)=V_{\text{DC}} + V_{\text{AC}}\sin{(2\pi f_a t)}$. Here, $V_{\text{DC}}$ represents the DC bias voltage, and $V_{\text{AC}}$ is the amplitude of the time-harmonic AC component, set at 0.5 V.

During the frequency up-sweep, the frequency $f_a$ increases linearly with time, ranging from 30 to 120 kHz. We determined the natural frequency of the beam's vibration, denoted as $f_n$, for different equilibrium deflections determined by the DC bias voltage. The measured resonant frequency of the beam, when zero electrostatic voltage is applied, is approximately 111 kHz. 

\begin{figure}[!ht]
\centering
\includegraphics[width=0.75\textwidth]{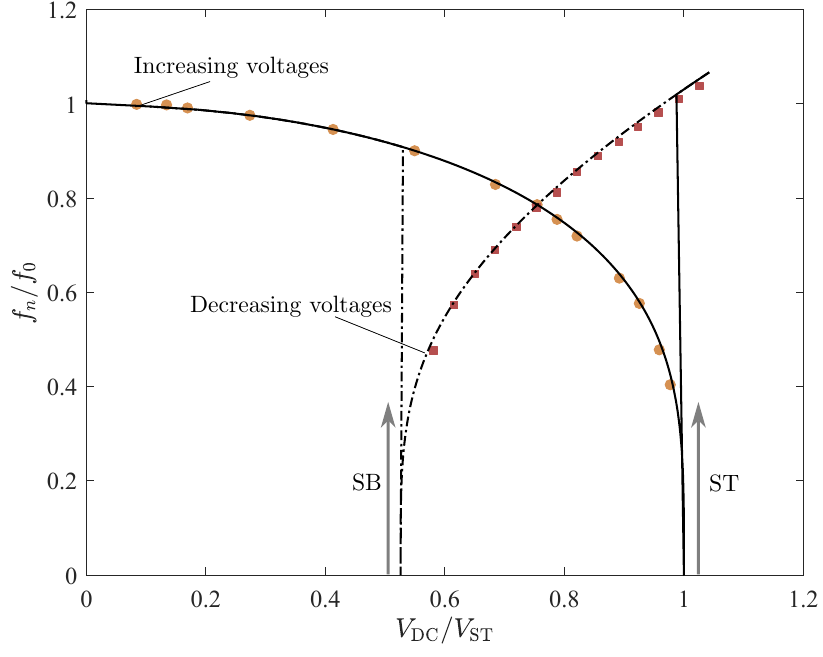}
\caption{Frequency of free vibrations around equilibrium as a function of the increasing/decreasing DC bias voltage. The baseline value $f_{0}$ represents the structure's frequency in the initial unloaded configuration. The snap-through voltage $V_{\mathrm{ST}}$ normalized the bias voltage. 
Experimental results for increasing and decreasing voltages are represented by circle and square markers, respectively. The solid and dash-dot lines represent the calculated (model) increasing/decreasing curves. 
\label{fig:fv} }
\end{figure}

In Figure~\ref{fig:fv}, we present the normalized resonant frequency of the beam as a function of the normalized bias voltage. 
The experimental data is depicted as circles and squares, while the calculated results, obtained using Eq.~\eqref{eq:NF}, are represented by solid and dash-dot lines. These dependencies are displayed for increasing and decreasing DC voltage ($V_{\text{DC}}$) values, covering the range encompassing the values at which ST and SB events (represented by $V_\text{ST}$ and $V_\text{SB}$) occur. Notably, the experimental findings align well with the results generated by the reduced-order model.

As expected, the vibration frequency decreases near the ST and SB points. However, it is crucial to note that the experimental data does not reach absolute zero values, which would entail zero frequency and stiffness. This contrasts with the model's predictions~\cite{krylov2010dynamic}.

The measured frequencies at the ST and SB branches reveal an exceptionally low beam stiffness near the critical \VST~and \VSB~points. Consequently, even a minor vibration and noise level can induce an ST transition at voltages lower than initially predicted.


We can derive consistent conclusions by examining the beam's frequency response in electrostatic actuation with a linear voltage ramp. In this analysis, we utilized the same voltage ramp signal, denoted as \VES, as previously detailed in Section~\ref{sec:exp_methods}, with a 60-second duration, allowing for a thorough examination of the beam's vibrations. The LDV signal was sampled at a rate of 1 MHz.

\begin{figure}[!ht]
\centering
\includegraphics[width=0.75\textwidth]{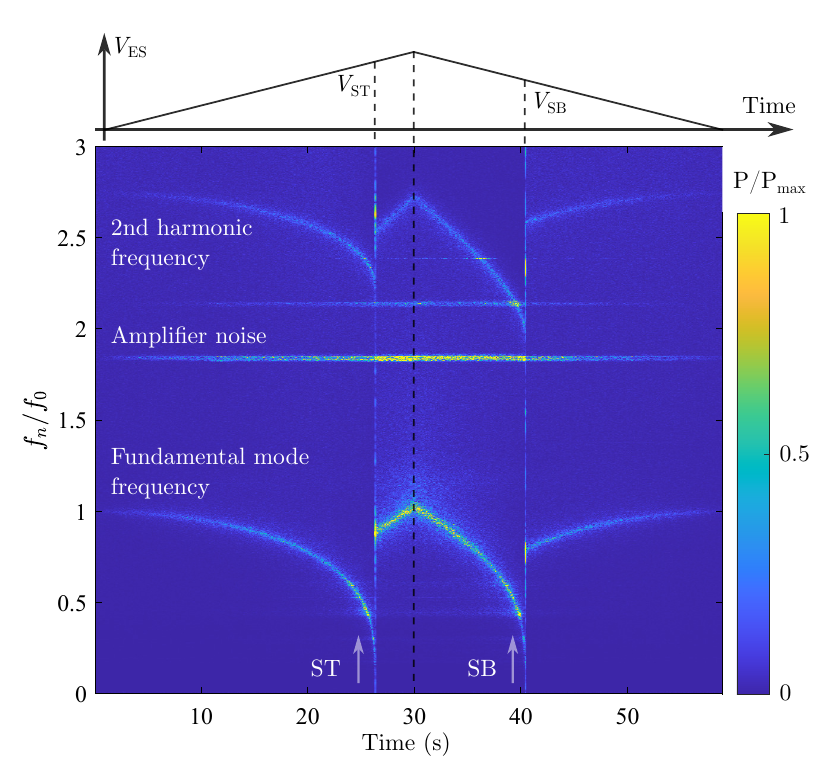}
\caption{Frequency response of the beam vibrations in the form of a spectrogram during the voltage actuation ramp \VES~(60 s). The level of the PSD is normalized to the maximum value. The overheat ratio is $V_{\mathrm{h}}=0$ V. The line corresponding to $1.8f_0$ is a regular noise signal from the voltage amplifier.
\label{fig:spectrogram}}
\end{figure}

Figure~\ref{fig:spectrogram} illustrates the beam's frequency response, particularly during the occurrences of snap-through (ST) and snap-back (SB) transitions. This power spectral density (PSD) spectrogram was generated through the application of the overlapped segment averaging spectral estimation (using the 'spectrogram' Matlab function) with a segment length of $2^{18}$ samples and a 75\% overlap.

The spectrogram reveals that the fundamental frequency closely mirrors the curves in Figure~\ref{fig:fv}. Notably, a sharp change in the fundamental frequency is evident during the transition between equilibrium states, specifically during snap-through and snap-back. Moreover, we observe the emergence of a higher (second) harmonic vibration mode, surpassing the intensity of the first mode by a factor of 2.75. These results affirm our findings regarding the sensor's high sensitivity in proximity to the beam's equilibrium state. They underscore that even a minor level of vibration or noise can trigger snap-through or snap-back transitions. 
These factors are known to hinder the experimental voltage-frequency dependency curves from reaching the theoretical zero level, as documented in references~\cite{tilmans1994electrostatically, hajjaj2017scalable, torteman2019micro}.


To gain a deeper insight into this effect, we performed an experiment designed to emulate the influence of the noise on the ST and the SB voltages. Namely, we add an electrical white noise signal generated by the wave generator to the baseline DC electrostatic voltage \VES~used in all other experiments. 

\begin{figure}[!ht]
\centering
\includegraphics[width=1\textwidth]{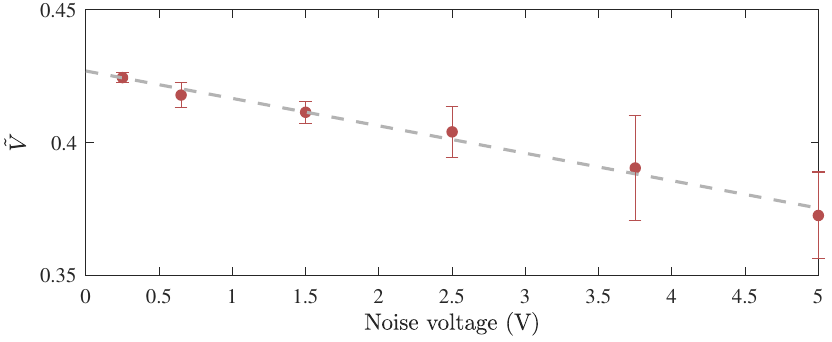}

\caption{Effect of noise on sensor performance (\VET=0 V). The error bars represent a standard error of the mean.
\label{fig:NoiseVSTVSB} }
\end{figure}

The noise amplitude, ranging from 0 to 5 V, should be viewed in the context of the base DC voltage, which extends from 0 to 200 V in the electrostatic voltage ramp signal. The measurements of \VST~and \VSB~are carried out under conditions without flow ($U=0$ m/s) and are depicted in Figure~\ref{fig:NoiseVSTVSB}.
As the amplitude of noise increases, its influence becomes apparent in the absolute value of $\tilde{V}$. This provides direct evidence that the beam exhibits remarkable sensitivity to fluctuations, including those induced by the flow, near the critical equilibrium point. Consequently, it is well-suited for sensing turbulent flows. This capability has the potential for applications beyond the measurement of average flow velocity. It can also capture turbulent fluctuations, such as in shear stress measurements, particularly when the sensors are flush with walls. Notably, similar sensors based on hot-wire technology have been developed~\cite{chew1998dynamic, sturzebecher2001surface, ghouila2019unsteady}. Our findings open up new possibilities for applying bifurcation-based sensors in this field. Note that since LDV sensing is suitable only in a laboratory environment, snap-through events detection based on electrostatic, piezoelectric, or piezoresistive sensing should be developed in the future to be implemented in integrated packaged sensors used in field experiments.


\subsection{Dynamics of the ST transition}



In contrast to previous studies~\cite{kessler2020flow, kessler2021sampling}, where a single-point LDV system was used to measure only the deflection at the midpoint, our approach in this study takes a significant step forward. We utilize the Micro System Analyzer (Polytec, MSA 600).

The MSA operates in scanning mode, allowing us to capture the complete shape of the entire beam during the transient ST and SB events. To achieve this, the MSA records a 36 milliseconds long velocity signal from 56 points distributed along the length of the beam. We actuate the beam buckling with a linear ramp of $V_{\mathrm{ES}}$ ranging from 0 to 200 V at a frequency of 400 Hz. The displacement of the beam at each point is determined by integrating the velocity signal over time. Phase averaging is applied to calculate the average displacement at each beam measurement point, eliminating free vibrations after ST and SB.

\begin{figure}[!ht]
\centering
\includegraphics[width=1\textwidth]{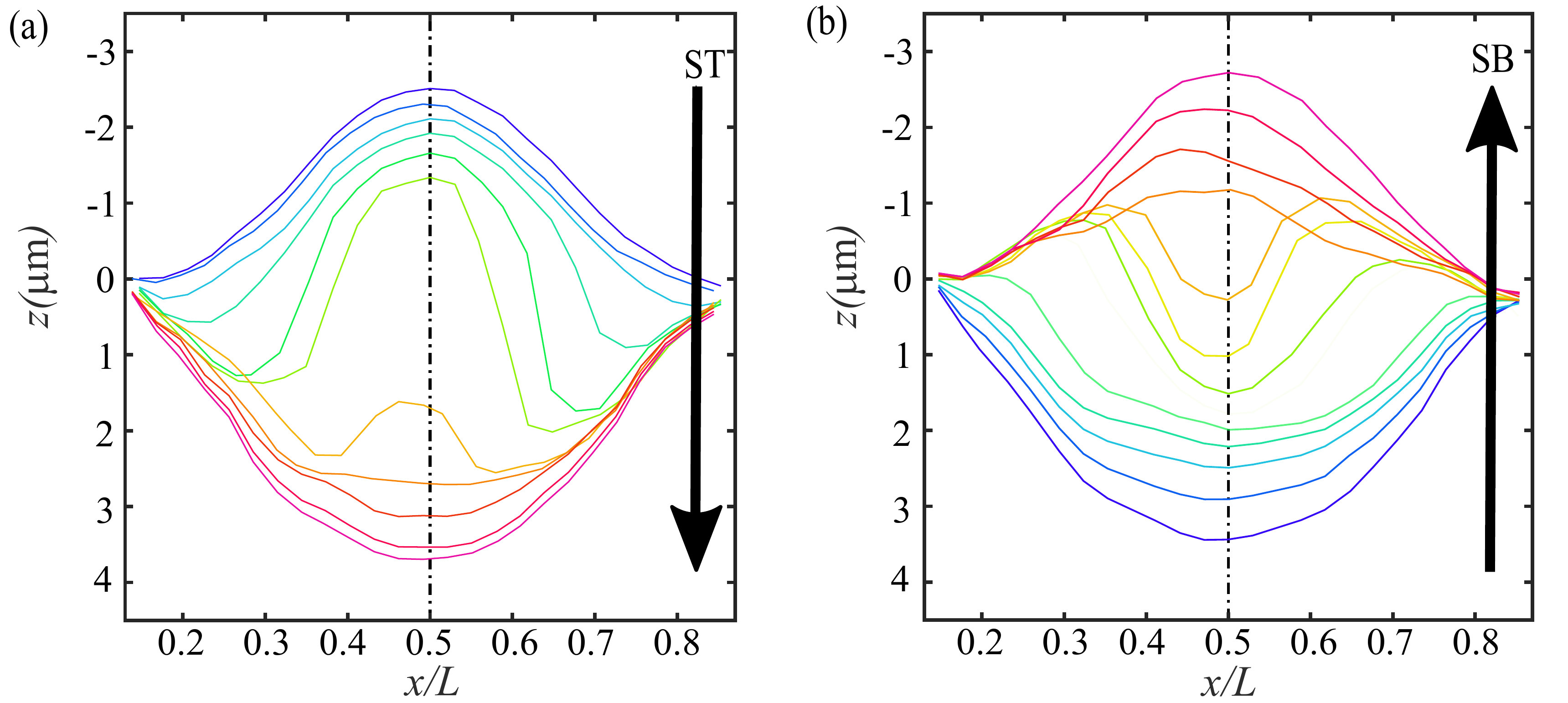}
\caption{Snapshots of beam deformation during snap-through transition (a), and snap-back transition (b).
\label{fig:snapshots} }
\end{figure}

In Figure~\ref{fig:snapshots}, we present snapshots of the beam's shape $z(x)$ along the beam's length $x$ during the snap-through and snap-back transitions. 
The arrows indicate the direction of the transitions, and the color coding reflects the progression from blue to magenta over time. We have marked positions at various arbitrary times, evenly distributed for clarity.

This represents a real-time beam profile measurement during the dynamic snap-through and snap-back transitions. In particular, we observe a slight asymmetry in the shape of $z(x,t)$ during both transitions concerning the midpoint, $z(x=0.5 L)$. This implies that the transitions occur at lower \VST~and higher \VSB~values than what would be anticipated from a symmetric first-mode transition. The presentation of these asymmetric snapshots of beam deformation during ST and SB events is a novel contribution, supporting the hypothesis presented in~\cite{medina2012symmetry} regarding the nature of this transition. Additionally, it provides insight into the shape of free vibrations following snap-through, in line with findings from~\cite{song2021free}.

Notably, during the beam transition, it is evident that the center of the beam lags behind the rest of the beam, except the edges, which remain static. Notably, the sections between the center and the edges, around $0.2<x/L<0.4$ and $0.6<x/L<0.8$, lead the motion during a significant portion of the deformation. This shape likely reflects a combination of various modes and underscores the contribution of the second symmetric mode. This discovery encourages further exploration in this direction.


\section{Summary and conclusions}
\label{sec:conclusion}

This study delved into critical aspects of the sensor's behavior. First, we quantified the ``loading effect'' induced by the laser Doppler vibrometer, which sheds light on the impact of laser heating on voltage measurements.

Second, we investigated the beam's response to varying levels of Joule heating. While the heating voltage \VET~influenced the absolute values of \VST~and \VSB~voltages, the device's sensitivity remained notably consistent. Our analytical model supported these findings and revealed a latching effect with increasing Joule heating. These results suggest that the beam overheat ratio may not be as pivotal as assumed for most thermal MEMS sensors, promising substantial energy consumption reduction.

Third, we investigate the influence of the flow angle on the beam's performance. We conclude that the device performs similarly to previous hot-wire-based microsensors and can measure flow direction.

We analyzed in detail the beam's response to fluctuations at different frequencies, typically present in various applications. Our results confirmed that this device, operating near critical transition points, is also sensitive to turbulent flow velocity fluctuations.

Finally, we obtained the first direct measurements of the beam's shape during transitions and confirmed the theoretical predictions of asymmetric, higher-than-first-mode transitions. We can confirm that the beam's single-degree-of-freedom approximation needs further improvement.

Overall, this study lays the groundwork for designing a series of flow sensors and multisensor arrays for velocity and velocity gradient measurements using initially curved bistable microbeams. However, unanswered questions still require further investigation, such as the sensor's response to out-of-plane flow directions, its performance in realistic environments, and the noise levels at which it can accurately measure low- and high-flow velocities for different applications.

\section*{Acknowledgments}

The authors acknowledge the financial support of the Israeli Innovation Authority, NOFAR program, and RAFAEL LTD, Advanced Defense Systems. S. Krylov acknowledges the support of the Henry and Dinah Krongold Chair of Microelectronics. The authors would also like to thank Gal Spaer Milo,
Grigori Gulitski and Stella Lulinsky for their assistance.



 \bibliographystyle{elsarticle-num} 
 \bibliography{references}





\end{document}